# Effect of AWGN & Fading (Raleigh & Rician) channels on BER performance of a WiMAX communication System

*Nuzhat Tasneem Awon*
Dept. of Information & Communication Engineering
University of Rajshahi, Rajshahi, Bangladesh
e-mail: tasneemawon@gmail.com

*Md. Mizanur Rahman*
Dept. of Information & Communication Engineering
University of Rajshahi, Rajshahi, Bangladesh
e-mail: mizan5624@yahoo.com

*Md. Ashraful Islam*
Lecturer
Dept. of Information & Communication Engineering
University of Rajshahi, Rajshahi, Bangladesh
e-mail: ras5615@gmail.com

*A.Z.M. Touhidul Islam*
Associate Professor
Dept. of Information & Communication Engineering
University of Rajshahi, Rajshahi, Bangladesh
e-mail: touhid_ict_it@yahoo.com

*Abstract*— The emergence of WIMAX has attracted significant interests from all fields of wireless communications including students, researchers, system engineers and operators. The WIMAX can also be considered to be the main technology in the implementation of other networks like wireless sensor networks. Developing an understanding of the WIMAX system can be achieved by looking at the model of the WIMAX system. This paper discusses the model building of the WIMAX physical layer using computer MATLAB 7.5 versions. This model is a useful tool for BER (Bit error rate) performance evaluation for the real data communication by the WIMAX physical layer under different communication channels AWGN and fading channel (Rayleigh and Rician), different channel encoding rates and digital modulation schemes which is described in this paper. This paper investigates the effect of communication channels of IEEE 802.16 OFDM based WIMAX Physical Layer. The performance measures we presented in this paper are: the bit error rate (BER) versus the ratio of bit energy to noise power spectral density (Eb/No). The system parameters used in this paper are based on IEEE 802.16 standards. The simulation model built for this research work, demonstrates that AWGN channel has better performance than Rayleigh and Rician fading channels. Synthetic data is used to simulate this research work.

I. INTRODUCTION

The wireless broadband technologies are bringing the broadband experience closes to a wireless context to their subscribers by providing certain features, convenience and unique benefits. These broadband services can be categorized into two types; Fixed Wireless Broadband and Mobile Broadband. The fixed wireless broadband provides services that are similar to the services offered by the fixed line broadband. But wireless medium is used for fixed wireless broadband and that is their only difference. The mobile broadband offers broadband services with an addition namely the concept of mobility and nomadicity. The term nomadicity can be defined as "Ability to establish the connection with the network from different locations via different base stations" while mobility is "the ability to keep ongoing connections engaged and active while moving at vehicular speeds". Examples of wireless broadband technologies are Wireless LAN and WIMAX.

WIMAX is the abbreviation of Worldwide Interoperability for Microwave Access and is based on Wireless Metropolitan Area Networking (WMAN). The WMAN standard has been developed by the IEEE 802.16 group which is also adopted by European Telecommunication Standard Institute (ETSI) in High Performance Radio Metropolitan Area Network, i.e., the HiperMAN group. The main purpose of WIMAX is to provide broadband facilities by using wireless communication [1]. WIMAX is also known as "Last Mile" broadband wireless access technology WIMAX gives an alternate and better solution compared to cable, DSL and Wi-Fi technologies as depicted in Figure-a: [2]

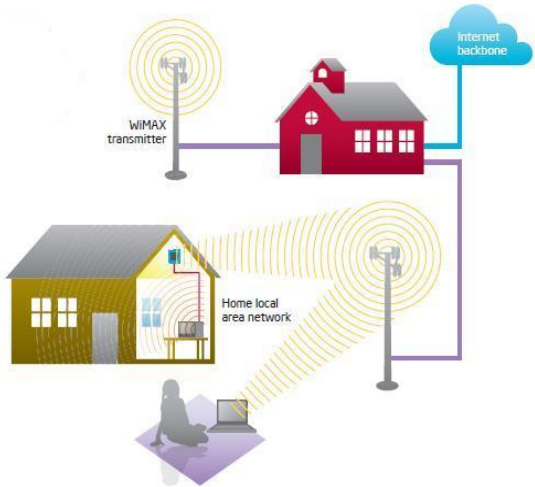

Figure-a: WiMAX System

Like other wireless communication network, transmission medium faces two major problems in WIMAX communication system. These problems are:
a) AWGN noise &
b) Rayleigh and Rician Fading.

**AWGN noise**
AWGN is a noise that affects the transmitted signal when it passes through the channel. It contains a uniform continuous frequency spectrum over a particular frequency band.

**Rayleigh Fading**
When no LOS path exists in between transmitter and receiver, but only have indirect path than the resultant signal received at the receiver will be the sum of all the reflected and scattered waves.

**Rician Fading**
It occurs when there is a LOS as well as the non-LOS path in between the transmitter and receiver, i.e. the received signal comprises on both the direct and scattered multipath waves. [2]

The objective of this project is to implement and simulate the IEEE 802.16 OFDM based WiMAX Physical Layer using MATLAB in order to have a better understanding of the standards and evaluate the system performance based on the effect of different communication channels. This involves studying through simulation, the various PHY modulations, coding schemes and evaluating the bit error rate (BER) performance of the WIMAX communication system under different channel models such as, AWGN channel and Fading (Rayleigh & Rician) channels.

II. SIMULATION MODEL

The transmitter and receiver sections of the WiMAX Physical layer are shown in the block diagram of Figure-b. This structure corresponds to the physical layer of the WiMAX air interface. In this setup, we have just implemented the mandatory features of the specification, while leaving the implementation of optional features for future work. The channel coding part is composed of coding techniques of the Cyclic Redundancy Check (CRC) and Convolutional Code (CC). The complementary operations are applied in the reverse order at channel decoding in the receiver end. We do not explain each block in details. Here we only give the emphasis on communication channel i.e. AWGN and Fading (Rayleigh and Rician) and Cyclic Redundancy Check (CRC) and Convolutional Code (CC) coding techniques.

A Convolution encoder consists of a shift register which provides temporary storage and a shifting operation for the input bits and exclusive-OR logic circuits which generate the coded output from the bits currently held in the shift register. In general, k data bits may be shifted into the register at once, and n code bits generated. In practice, it is often the case that $k$=1 and $n$=2, giving rise to a rate 1/2 code [3].

Cyclic Redundancy Check (CRC) codes are a subset of then class of linear codes, which satisfy the cyclic shift property such as if $C=[C_{n-1}, C_{n-2} ……, C_O]$ is a codeword of a cyclic code, then $[C_{n-2}\ C_{n-2}, …, C_0, C_{n-1}]$, obtained by a cyclic shifts of the elements of C, is also a code word. In other word all cyclic shifts of C are code words. From the cyclic property, the codes possess a great deal of structure which is exploited to greatly simplify the encoding and decoding operation [4].

Reasonable assumption for a fixed, LOS wireless channel is the additive white Gaussian noise (AWGN) channel [5], which is flat and not "frequency-selective" as in the case of the fading channel. Particularly fast, deep frequency-selective fading as often observed in mobile communications is not considered in this thesis, since the transmitter and receiver are both fixed. This type of channel delays the signal and corrupts it with AWGN. The AWGN is assumed to have a constant PSD over the channel bandwidth, and a Gaussian amplitude probability density function. This Gaussian noise is added to the transmitted signal prior to the reception at the receiver as shown in Figure-c [6], therefore the transmitted signal, white Gaussian noise and received signal are expressed by the following equation with s(t), n(t) and r(t) representing those signals respectively:

$$r(t)=s(t)+n(t)$$

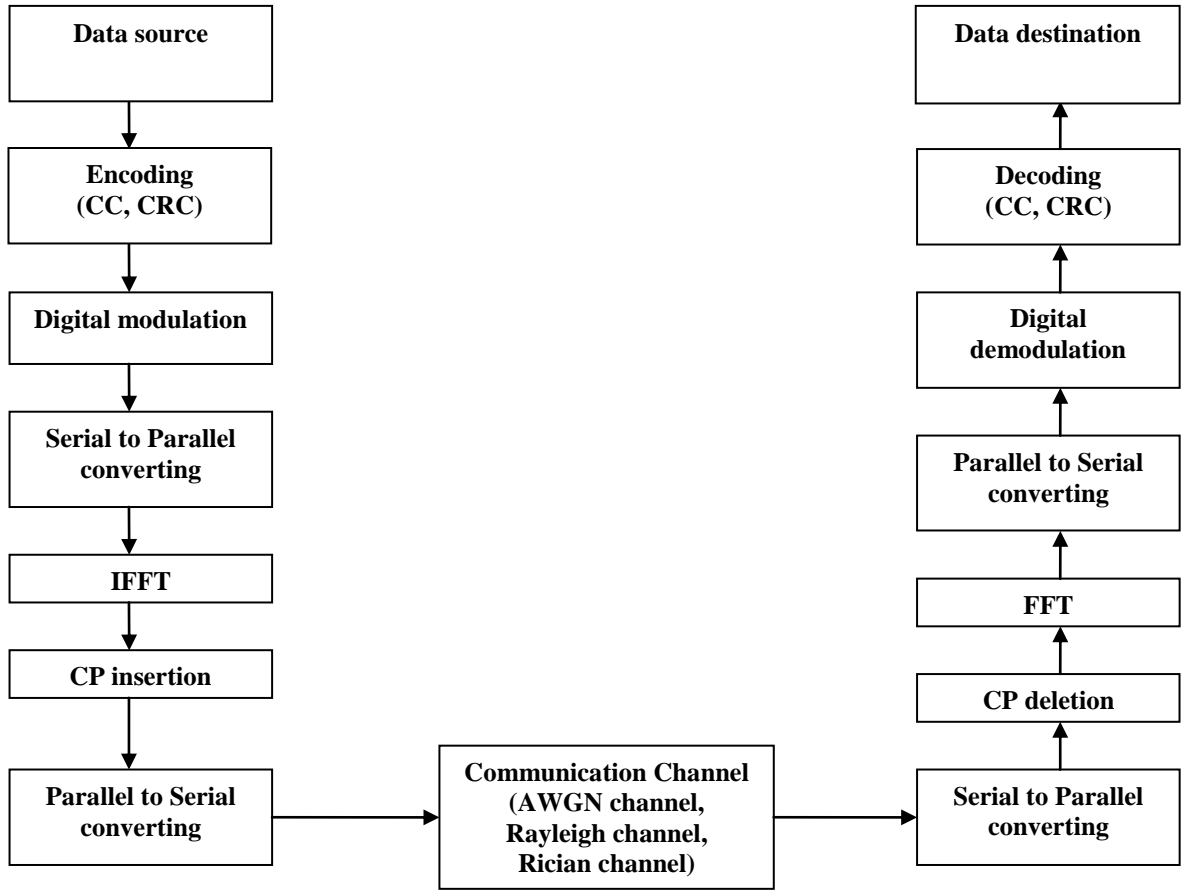

Figure-b: A block diagram for WIMAX Communication system

Where n(t) is a sample function of the AWGN process with probability density function (pdf) and power spectral density **[7].**

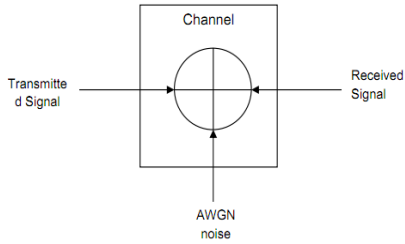

Figure-c: AWGN channel model

The in-phase and quadrature components of the AWGN are assumed to be statistically independent, stationary Gaussian noise process with zero mean and two-sided PSD of N0/2 Watts/Hz. As zero-mean Gaussian noise is completely characterized by its variance, this model is particularly simple to use in the detection of signals and in the design of optimum receivers [6]. So, it was developed using 'awgn' function which is also available in Matlab.

Multipath fading results in fluctuations of the signal amplitude because of the addition of signals arriving with different phases. This phase difference is caused due to the fact that signals have traveled different distances by traveling along different paths. Because the phases of the arriving paths are changing rapidly, the received signal amplitude undergoes rapid fluctuation that is often modeled as a random variable with a particular distribution.

The most commonly used distribution for multipath fast fading is the Rayleigh distribution, whose probability density function (pdf) is given by

$$f_{ray}(r) = \frac{r}{\sigma^2} \exp\left(-\frac{r^2}{2\sigma^2}\right), \quad r \geq 0$$

Here, it is assumed that all signals suffer nearly the same attenuation, but arrive with different phases. The random variable corresponding to the signal amplitude is r. Here $\sigma^2$ is the variance of the in-phase and quadrature components. Theoretical considerations indicate that the sum of such signals will result in the amplitude having the Rayleigh distribution of the above equation. This is also supported by measurements at various frequencies. The phase of the complex envelope of the received signal is normally assumed to be uniformly distributed in $[0, 2\pi]$.

When strong LOS signal components also exist, the distribution is found to be Rician, the pdf of such function is given by:

$$f_{ric}(r) = \frac{r}{\sigma^2} \exp\left(\frac{-(r^2 + A^2)}{2\sigma^2}\right) I_0\left(\frac{Ar}{\sigma^2}\right), \quad r \geq 0, A \geq 0$$

Where $\sigma^2$ is the variance of the in-phase and quadrature components. A is the amplitude of the signal of the dominant path and I0 is the zero-order modified Bessel function of the first kind. Normally the dominant path significantly reduces the depth of fading, and in terms of BER Ricean fading provides superior performance to Rayleigh fading. The probability of having line-of-sight (LOS) component depends on the size of the cell. The smaller the cell the higher the probability of having LOS path. If there is no dominant path then the Rician pdf reduces to Rayleigh pdf. When A is large compared with σ, the distribution is approximately Gaussian. Thus, since Ricean distribution covers also Gaussian and Rayleigh distribution, mathematically the Ricean fading channel can be considered to be general case [8].

The procedures that we have followed to develop the WiMAX physical layer simulator is briefly stated as follows:
At the transmission section:

1. At first we have generated a random data stream of length 44000 bit as our input binary data using Matlab 7.5. Then randomization process has been carried out to scramble the data in order to convert long sequences of 0's or 1's in a random sequence to improve the coding performance.
2. Secondly we have performed Cyclic Redundancy Check (CRC) encoding. After this 1/2 rated convolutional encoding is also implemented on the CRC encoded data. The encoding section was completed by interleaving the encoded data.
3. Then various digital modulation techniques, as specified in WiMAX Physical layer namely QAM, 16-QAM and 64-QAM are used to modulate the encoded data.
4. The modulated data in the frequency domain is then converted into time domain data by performing IFFT on it.
5. For reducing inter-symbol interference (ISI) cyclic prefix has been added with the time domain data.
6. Finally the modulated parallel data were converted into serial data stream and transmitted through different communication channels.
7. Using Matlab built-in functions, "awgn", "rayleighchan" and "ricianchan" we have generated AWGN, Rayleigh and Rician channels respectively.

At the receiving section we have just reversed the procedures that we have performed at the transmission section. After ensuring that the WiMAX PHY layer simulator is working properly we started to evaluate the performance of our developed system. For this purpose we have varied encoding

14

techniques and digital modulation schemes under AWGN and frequency-flat fading (Rayleigh/ Rician) channels. Bit Error Rate (BER) calculation against different Signal-to-Noise ratio (SNR) was adopted to evaluate the performance.

The simulation Parameters used in the present study are shown in Table 1.

**Table 1: Simulation Parameters**

| Parameters | values |
|---|---|
| Number Of Bits | 44000 |
| Number Of Subscribers | 200 |
| FFT Size | 256 |
| CP | 1/4 |
| Coding | Convolutional Coding(CC), Cyclic redundancy Check (CRC) |
| Code rate | CC(1/2) ,CRC(2/3) |
| Constraint length | 7 |
| K-factor | 3 |
| Maximum Doppler shift | 100/40Hz |
| SNR | 0-30 |
| Modulation | QAM, 16-QAM, 64-QAM |
| Noise Channels | AWGN, Rayleigh and Rician |

### III. SIMULATION RESULT

This section of the chapter presents and discusses all of the results obtained by the computer simulation program written in Matlab7.5, following the analytical approach of a wireless communication system considering AWGN, Rayleigh Fading and Rician Fading channel. A test case is considered with the synthetically generated data. The results are represented in terms of bit energy to noise power spectral density ratio (Eb/No) and bit error rate (BER) for practical values of system parameters.

By varying SNR, the plot of Eb/No vs. BER was drawn with the help of "semilogy" function. The Bit Error Rate (BER) plot obtained in the performance analysis showed that model works well on Signal to Noise Ratio (SNR) less than 25 dB. Simulation results in figure 5.1, figure 5.2 and figure 5.3 shows the performance of the system over AWGN and fading (Rayleigh & Rician) channels using QAM, 16-QAM and 64-QAM modulation schemes respectively.

**Performance of OFDM based WIMAX Physical layer using QAM modulation technique:**
The following figure shows the BER performance of WIMAX Physical layer through AWGN channel, Rayleigh and Rician fading channels using Quadrature Amplitude Modulation (QAM) technique. The effect of AWGN channel and fading (Rayleigh & Rician) channels, we get through this figure has been discussed later.

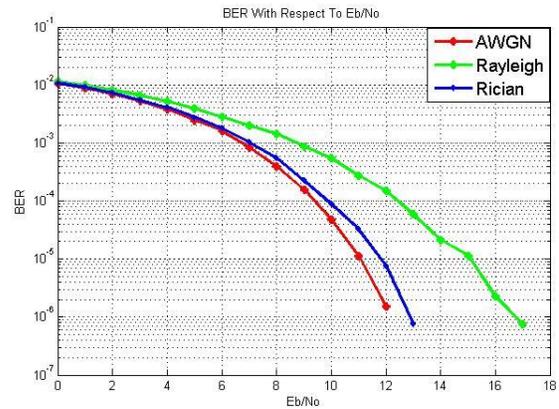

Figure-d: Bit error rate (BER) performance of AWGN, Raleigh and Rician channels for QAM modulation technique.

**Performance of OFDM based WIMAX Physical layer using 16-QAM modulation technique:**
The following figure shows the BER performance of WIMAX Physical layer through AWGN channel, Rayleigh and Rician fading channels using 16-QAM technique. The effect of AWGN channel and fading (Rayleigh & Rician) channels, we get through this figure has been discussed later.

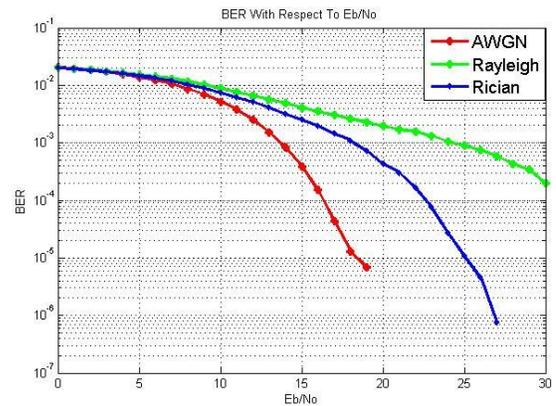

Figure-e: Bit error rate (BER) performance of AWGN, Raleigh and Rician channels for 16-QAM modulation technique.

**Performance of OFDM based WIMAX Physical layer using 64-QAM modulation technique:**

The following figure shows the BER performance of WIMAX Physical layer through AWGN channel, Rayleigh and Rician fading channels using 64-QAM technique. The effect of AWGN channel and fading (Rayleigh & Rician) channels, we get through this figure has been discussed later.

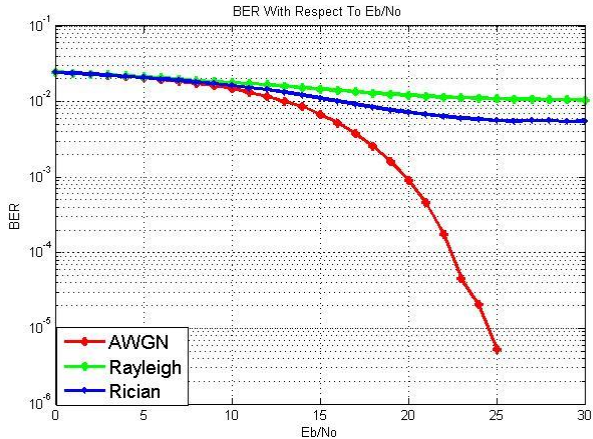

Figure-f: Bit error rate (BER) performance of AWGN, Raleigh and Rician channels for 64-QAM modulation technique.

**Effect of AWGN channel on BER performance of WIMAX Physical layer:**

From figure-d, e & f, we can see that, AWGN channel has lower BER than Raleigh and Rician fading channel. For an example, while using the QAM modulation scheme, for SNR value 13, BER for AWGN channel remains 0, where BER for Rayleigh and Rician channel remains 5.9091e-05 and 7.5758e-07 respectively. After SNR value 13, BER for AWGN remains zero for the rest of the SNR values. But Raleigh & Rician fading channel has more non-zero BER values than that of AWGN channel

**Effect of Raleigh fading channel on BER performance of WIMAX Physical layer:**

From figure d, e & f, we can see that, Raleigh fading channel has higher BER than AWGN and Rician fading channel. For an example, while using the QAM modulation scheme, for SNR value 17, BER for Raleigh fading channel remains 7.5758e-07, where BER for both AWGN and Rician channel remains zero. After SNR value 12 and after SNR value 13, BER for AWGN and for Rician fading channel remains zero for the rest of the SNR values, where Rayleigh fading channel has more non-zero BER values.

**Effect of Rician fading channel on BER performance of WIMAX Physical layer:**

From figure d, e & f, we can see that, Rician fading channel has higher bit error rate (BER) than AWGN channel, but lower than that of Rayleigh fading channel. For an example, while using the QAM modulation scheme, for SNR value 13, BER for Rician fading channel remains 7.5758e-07 while AWGN channel has zero BER. Again, for SNR value 14, both Rician fading channel and AWGN channel has zero BER while Rayleigh fading channel has BER value 2.1212e-05. After that, for SNR value 15 to 17, BER for Raleigh fading channel remains non-zero while BER for AWGN & Rician fading channels remain zero.

IV. CONCLUTION

In this research work, it has been studied the performance of an OFDM based WIMAX Communication system adopting different coding schemes and digital modulation scheme; M-ary QAM. A range of system performance results highlights the impact of AWGN and fading (Rayleigh & Rician) channels under QAM, 16-QAM & 64-QAM modulation techniques. From this research work, conclusions can be drawn regarding the BER performance evaluation of WIMAX Communication system over AWGN channel and fading (Rayleigh & Rician) channels like as below:

**1.** The performance of AWGN channel is the best of all channels as it has the lowest bit error rate (BER) under QAM, 16-QAM & 64-QAM modulation schemes. The amount of noise occurs in the BER of this channel is quite slighter than fading channels.

**2.** The performance of Rayleigh fading channel is the worst of all channels as BER of this channel has been much affected by noise under QAM, 16-QAM & 64-QAM modulation schemes.

**3.** The performance of Rician fading channel is worse than that of AWGN channel and better than that of Rayleigh fading channel. Because Rician fading channel has higher BER than AWGN channel and lower than Rayleigh fading channel. BER of this channel has not been much affected by noise under QAM, 16-QAM & 64-QAM modulation schemes.